\documentclass[aps,twocolumn,letterpaper,prl,showpacs]{revtex4-1}
\usepackage{eurosym}
\usepackage{graphicx}
\usepackage{dcolumn}
\usepackage{bm}
\usepackage{color}
\usepackage{latexsym}
\usepackage{amssymb}

\begin{document}

\title{Effect of a marginal inclination on pattern formation \\
in a binary liquid mixture under thermal stress.}
\author{Fabrizio Croccolo$^{(1,2)}$, Frank Scheffold$^{(1)}$ and Alberto
Vailati$^{(3)}$}

\address{$^{(1)}$Department of Physics, University of Fribourg, CH-1700
Fribourg, Switzerland} 
\address{$^{(2)}$present address: Laboratoire des
Fluides Complexes et leurs R\'eservoirs - CNRS UMR5150, Universit\'e de Pau
et des Pays de l'Adour, 64600 Anglet, France} 
\address{$^{(3)}$ Dipartimento
di Fisica, Universit\'a degli Studi di Milano, I-20133 Milano, Italy}

\date{\today}

\begin{abstract}
Convective motions in a fluid layer are affected by its orientation with
respect to the gravitational field. We investigate the long-term stability
of a thermally-stressed layer of a binary liquid mixture and show that
pattern formation is strongly affected by marginal inclinations as small as
a few milliradians. At small Rayleigh numbers the mass transfer is dominated
by the induced large scale shear flow, while at larger Rayleigh numbers it
is dominated by solutal convection. At the transition, the balance between
the solutal and shear flows gives rise to drifting columnar flows moving in
opposite directions along parallel lanes in a Super-Highway configuration.

\end{abstract}

\pacs{47.20.Bp, 47.54.-r, 92.10.af}
\keywords{convective instabilities, soret effect, shadowgraph, fluid binary
mixture}
\maketitle


\bigskip

Pattern formation in non-equilibrium systems arises from symmetry breaking
of an isotropic initial state \cite{cross93}. Whenever two symmetry-breaking
mechanisms coexist their competition gives rise to a rich phase diagram. A
typical example is represented by an inclined layer of liquid under the
action of a temperature difference. Tilting the layer can determine a large
scale shear flow (LSF). For fairly large inclinations theoretical \cite%
{clever77, busse92, busse00} and experimental studies \cite{daniels00,
daniels02, seiden08, weiss12, zebib09} reveal rich spatio-temporal dynamics
in the phase diagram, characterized by the presence of Busse oscillations,
subharmonic oscillations as well as longitudinal and cross rolls. The
transitions between different regimes occur at angles of the order of tens
of degrees, and pattern formation does not appear to be influenced by small
inclination angles of the order of a few degrees or smaller.\ By adding a
second component to the mixture, the compositional stratification also
contributes to the convective behavior. A remarkable example is represented
by thermohaline circulation in oceans, where both the local salinity and
temperature of water contribute to the convective motions, the thickness of
the layer of water being modulated by the seafloor \cite{wunsch02,
rahmstorf03}. The thermal stress applied to a liquid mixture via a
temperature gradient can be quantified by the dimensionless Rayleigh number $%
Ra=g\Delta \rho _{T}h^{3}/(\eta D_{T})$ \cite{bookfaber95}, where $g$ is the
gravity acceleration, $\Delta \rho _{T}$ the density difference generated by
thermal dilation of the liquid, $h$ the liquid layer thickness, $\eta $ its
shear viscosity and $D_{T}$ the thermal diffusivity. The presence of the
thermal stress determines a separating flux of the two components and, in
turn, a solutal density difference $\Delta \rho _{c}$ throughout the sample,
whose value and orientation are quantified by the Soret coefficient $S_{T}$ 
\cite{soret79, bookdegroot62}. In the case of a mixture with positive $S_{T}$%
, as the one used in our study, both $\Delta \rho _{T}$ and $\Delta \rho _{c}
$ contribute to destabilize a sample heated from below. 
The relative importance of the two contributions is expressed by the separation ratio $%
\psi =\Delta \rho _{c}/\Delta \rho _{T}$. The concentration difference $%
\Delta \rho _{c}$ determines a solutal stress on the mixture expressed by
the solutal Rayleigh number $Ra_{s}=Ra\Psi /Le)$, where $Le=D/D_{T}$ is the
Lewis number of the mixture and $D$ its diffusion coefficient \cite{platten,
shevtsova06}. At Rayleigh numbers smaller than the threshold $Ra_{c}\simeq
1700$ for Rayleigh-B\'{e}nard convection the long persistence time of
concentration perturbations can give rise to sustained solutal convective
motions even in the presence of small $\Delta \rho _{c}$. \newline
\indent In this letter we investigate the influence of a small inclination
angle in the range $0$ mrad$<\alpha <36$ mrad ($\alpha <2^{\circ }$) on the
long-term convective behaviour of a layer of a binary liquid mixture under
thermal stress below the threshold for Rayleigh-B\'{e}nard convection. 
Interestingly, we find the long-term stability of the mixture to be strongly
influenced already by small inclination angles, in contrast to the
short-term behaviour which remains unaffected. The observed convective
patterns are arranged into a peculiar Super-Highway configuration,
reminiscent of the traffic of vehicles during rush hours. Such pattern
differs dramatically from the square, roll, and cross-roll patterns
previously reported experimentally \cite{legal85,moses86} and theoretically 
\cite{jung98,weggler10} for binary liquid mixtures heated from below in the
absence of inclination. \newline
\indent The binary mixture selected for this study is
isobutylbenzene/n-dodecane at 50\% weight fraction. The choice of this
mixture is motivated by the availability of an accurate and extensive set of
thermophysical properties \cite{properties}. The sample is delimited by two $%
8\times 40\times 40$ mm$^{3}$ square sapphire windows kept at fixed distance
and sealed by a circular Viton O-ring gasket with an inner radius $R=13$ mm.
The geometry corresponds to a moderately high aspect ratio $R/h=10$ of the
sample. The sapphire windows are in contact with two annular thermo electric
devices connected to two independent proportional-integral-derivative (PID)
temperature controllers. The temperature of each sapphire window can be
controlled independently with an absolute accuracy of $0.01$ K and a
relative RMS stability of $0.001$ K over 24 hours. The Peltier elements have
a central hole with radius of $r=6.5$ mm that determines a clear aperture
suitable to perform optical measurements. The performances and reliability
of the cell have been established during a long series of experiments on
devices sharing a similar conceptual design (see \cite%
{vailati11,bernardin12,croccolo12} and references therein). Calibration
measurements performed with water in strictly non-inclined conditions showed 
$Ra_{c}=1670\pm 50$. Finite element modeling has been performed to evaluate
the cell radial inhomogeneities of the temperature profile. We find the
inhomoegeneities to be smaller than $2\%$ of the vertical gradient,
comparable to the results previously reported \cite{vailati11}. The
uniformity of the cell thickness is better than $0.01mm$ over the entire
field of view. 

\begin{figure}[t]
\centering\includegraphics[width=8cm]{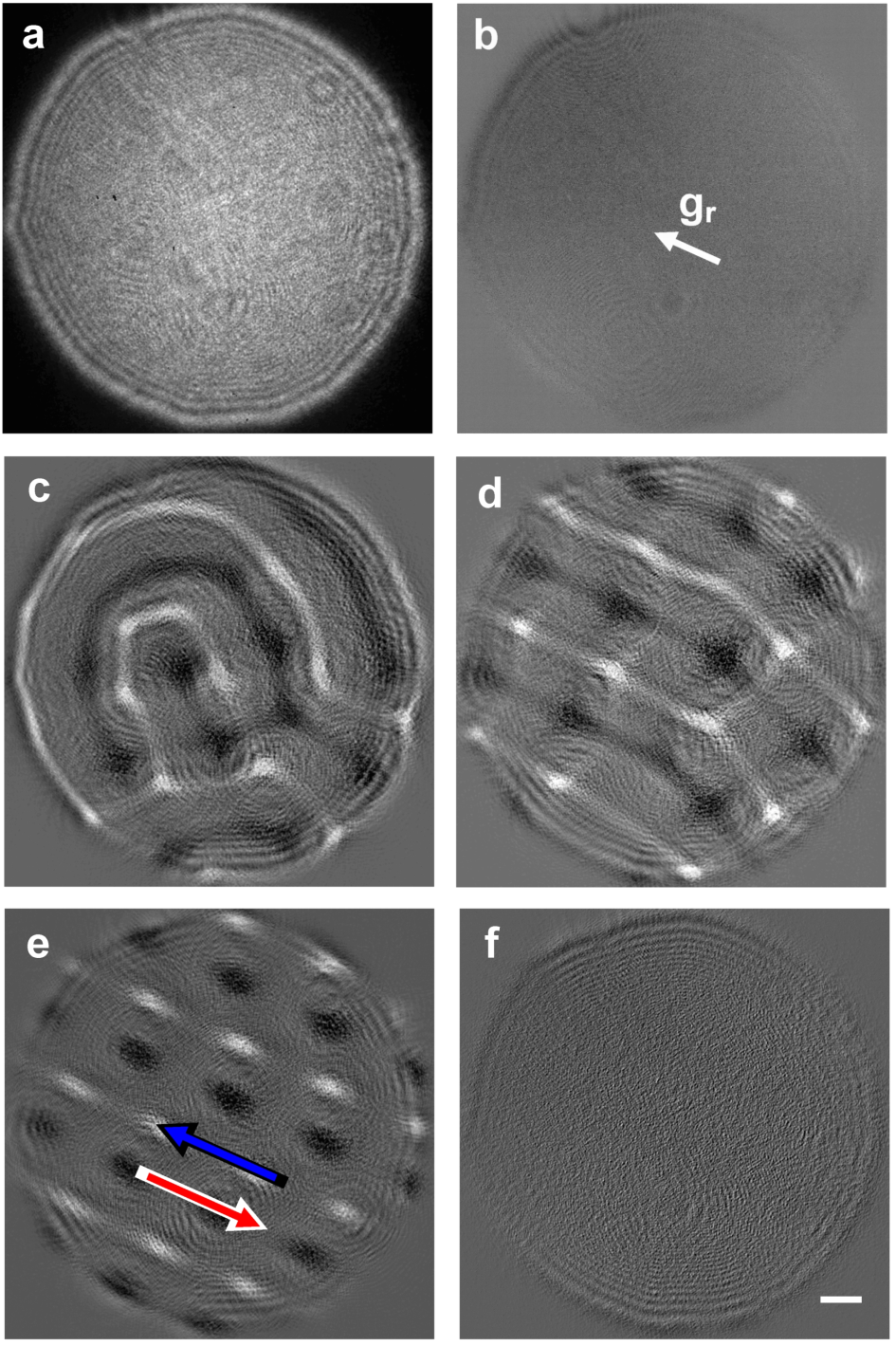}
\caption{(color online) Shadowgraph images of the convective patterns: $%
Ra=1320$, $\protect\alpha =24mrad$ a, Reference image taken at $t=0s$. b-f,
Difference images; dark zones denote the warmer fluid, which is less
concentrated in the denser isobutylbenzene component. Different times shown
are: b 150s, c 400s, d 4000s, e 20500s, and f 29000s. The direction of the
marginal gravitational acceleration parallel to the cell plane is indicated
by an arrow in b. The directions of motion of the lines for the
Super-Highway convection are indicated by red and blue arrows in e. The size
bar in f corresponds to the sample thickness $h=1.3mm$.}
\label{fig shad pics}
\end{figure}
A $h=1.30$ mm thick layer of the mixture is hosted inside the thermal
gradient cell that can be tilted by using a calibrated screw. We performed
careful control measurements at zero tilt angle, lasting as long as one
month. These measurements allowed us to exclude the presence of a large
scale flow when the cell is kept horizontal (see supplemental text). As a
result of these measurements we conservatively estimate that the accuracy of
the tilt angle is better than 2 mrad. The diagnostic method of choice is
shadowgraph, a visualization technique widely employed in fluid dynamics 
\cite{booksettles01, trainoff02, croccolo11}. Its implementation comprises a
super-luminous diode (Superlum, Broad Lighter S680) with a wavelength of $%
\lambda =(683\pm 9)$ nm, coupled to a single mode optical fiber as a light
source. The diverging beam out of the fiber is collimated by using an
achromatic doublet lens of focal length $f=150$ mm positioned at a focal
distance from the lens. No other lens is used after the sample cell. A
Charged Coupled Device sensor (Vossk\"{u}hler, CCD4000) with a resolution of
2048$\times $2048 pixels of $7.4\times 7.4$ $\mu $m$^{2}$ is placed at a
distance of $z=(260\pm 10)$ mm from the sample cell \cite{croccolo12}. 
\newline
\indent A typical measurement sequence involves the rapid imposition of a
temperature difference by heating from below. As a result, a nearly linear
temperature profile is established across the sample in a time $\tau
_{T}=h^{2}/D_{T}\cong 20$ s. During the process, we grab continuously
shadowgraph images of the convective patterns. The frame rate is set to 1Hz
during the initial fast kinetics and reduced to 1/60Hz for the subsequent
slower phase. A sampling from a typical image sequence corresponding to an
inclination of 24 mrad and to a Rayleigh number $Ra=1320$ ($\Delta T=8.5$ K)
is shown in Fig. 1 (see also Supplemental Movie S1). A reference background
image is taken before applying the temperature gradient to the fluid mixture
(Fig. 1a). This image is then subtracted to all the images subsequently
collected at generic time $t$. The imposition of the temperature difference
is followed by a latency time where no pattern formation occurs (Fig. 1b).
This featureless phase ends with the appearance of convective rolls (Fig.
1c). After a few thousands seconds the rolls rearrange towards a more
ordered quasi-squared pattern (Fig.1d), a configuration typical of solutal
convection in binary mixtures with positive Soret coefficient heated from
below \cite{legal85, moses86, jung98}. \newline
\indent Quite surprisingly, after a time of the order of $10^{4}$s the
square patterns start to drift and evolve gradually into sequences of
columns of liquid moving in opposite directions along parallel lanes
arranged in a Super-Highway configuration (SH) (Fig. 1e). Eventually, the SH
patterns might fade away and the convective pattern can become almost
featureless (Fig. 1f). At Rayleigh numbers of the order of $Ra_{SH}\approx
1400$ ($\Delta T=9$ K) the SH patterns become stable (Supplemental Movie
S2). At larger Rayleigh numbers, the patterns display marked oscillations
between a drifting square (DS) pattern (Supplemental Movie S3) and an SH
one. The long time behavior of the patterns is strongly affected by the
marginal inclination angle and by the Rayleigh number (Fig. 2). The phase
diagram displays a stable Large Scale Flow (LSF) region at small Rayleigh
numbers, where convective patterns are destroyed by the long-term effect of
the inclination. At higher Rayleigh numbers the columnar convection is more
efficient than the LSF. Under these conditions stable drifting square
patterns are present. The intermediate region is characterized by patterns
where the competition between LSF and solutal convection becomes apparent
and the system displays marked oscillations between drifting square patterns
and longitudinal rolls or SH patterns. In a narrow region of the phase space
(red stars in Fig.2) the balance between LSF and solutal convection gives
rise to a stable SH convection state in which the columns organize into
parallel lanes drifting in opposite directions (Fig.1e and Supplemental
Movie S4). The white spots, which represent descending columns of colder
liquid crossing the cell perpendicularly to the plane of the figure, align
into parallel lines and move in the direction of the residual gravitational
force $g_{r}$ parallel to cell surface, while the dark spots align into
another set of lines and move in the opposite direction. The two sets of
lines are staggered across the cell plane. \newline
\indent The different size of the two sets of spots is related to the
shadowgraph visualization because areas with a refractive index slightly
larger/smaller than the surroundings act like converging/diverging lenses 
\cite{trainoff02,croccolo11}. The development of lanes in the presence of two flows
occurring in opposite directions is a rather general self-organization
process that occurs also in other systems, such as groups of animals \cite%
{couzin}. In particular, simulations of the dynamics of groups of pedestrian
crossing a walkway in opposite directions show the development of lanes of
flow similar to those reported by us, but obtained in the presence of more
generic local interations than the hydrodynamic ones governing our system 
\cite{helbing}.

\begin{figure}[t]
\center\includegraphics[width=8cm]{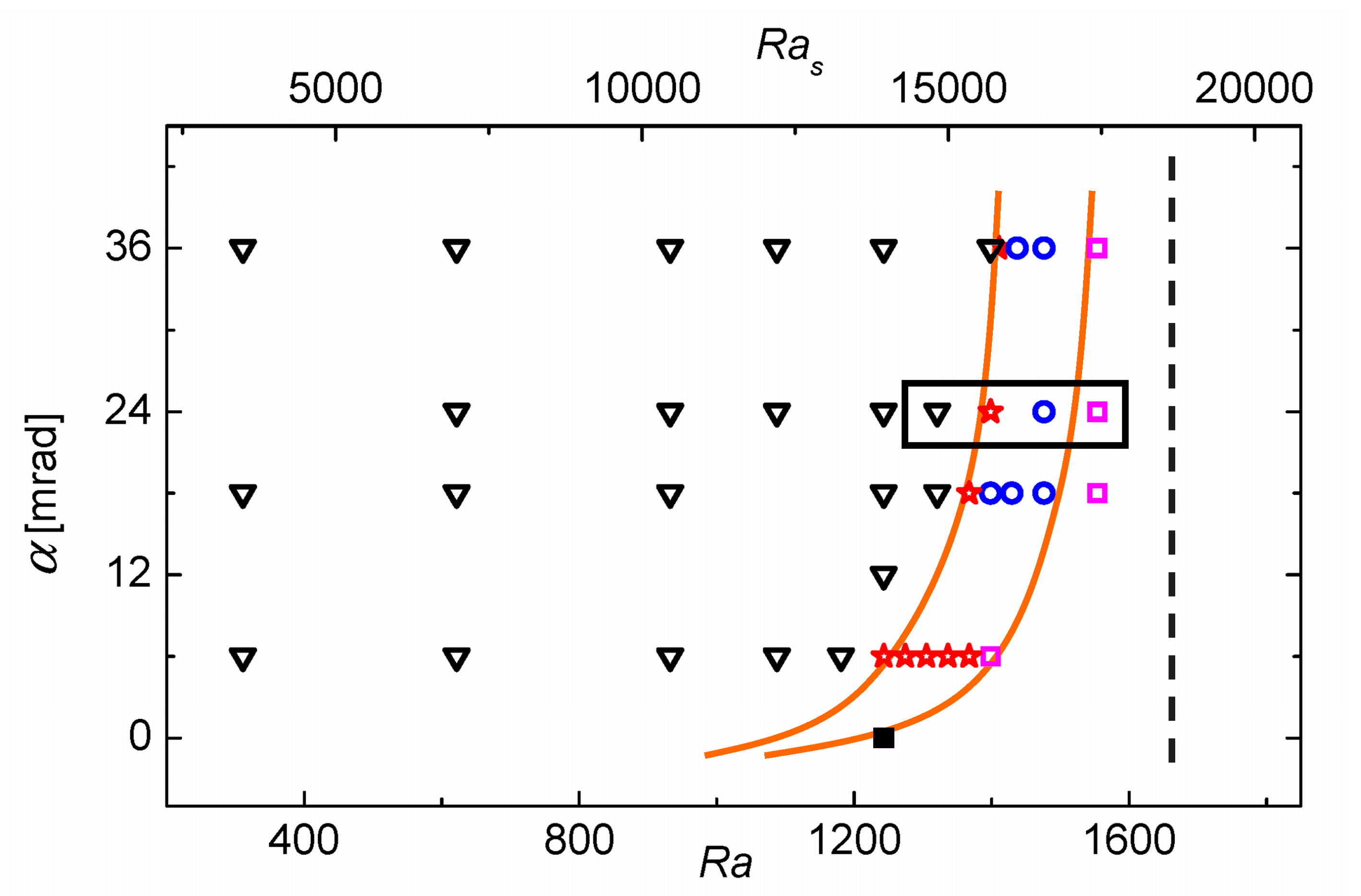}
\caption{(color online) Phase diagram of long-term convective behavior:
phase diagram of the convective pattern as a function of the Rayleigh number
(or solutal Rayleigh number,top axis) and the inclination angle. Symbols
represent the different long-term convective states: ($\triangledown $) for
Large Scale Flow (LSF);($\star $) for Super-Highway (SH) convection; ($\circ 
$) for oscillations; ($\square $) for drifting square patterns (DS); and ($
\blacksquare$) for stable still square patterns. The two curved lines mark the
approximate boundaries of the transition region, the vertical one indicates
the value of the critical Rayleigh number at zero inclination. The rectangle
outlines the experimental conditions shown in Fig. 3.}
\label{fig contrast}
\end{figure}
\indent In order to perform a systematic investigation of the evolution of
the convective patterns we determined from each image sequence the time
evolution of the contrast $C(t)$ of the images. The image contrast is
defined as $C(t)=<[i(t)-i(t_{o})]^{2}>_{\vec{x}}$ where $<...>_{\vec{x}}$
represents the average over the pixels of an image, $i(t)=I(t)/<I(t)>_{\vec{x%
}}$ is an image normalized by its spatial average, and $i(t_{o})$ is a
normalized background image at time $t_{o}$ before applying the temperature
gradient. The contrast provides a quantitative estimate of the strength of
the temperature and concentration modulations generated by the convective
motions. Under all the explored experimental conditions, after the
imposition of the temperature difference the contrast remains constant for a
latency time $t^{\ast }$ lasting from tens to hundreds of seconds (Fig. 3,
left arrows). The latency period is then followed by a rapid growth of the
contrast, and by relaxation oscillations leading to a value that remains
fairly stable for a second latency time $t_{LSF}$ related to the inclination
of the cell (Fig. 3, right arrows). 
\begin{figure}[b]
\center\includegraphics[width=8cm]{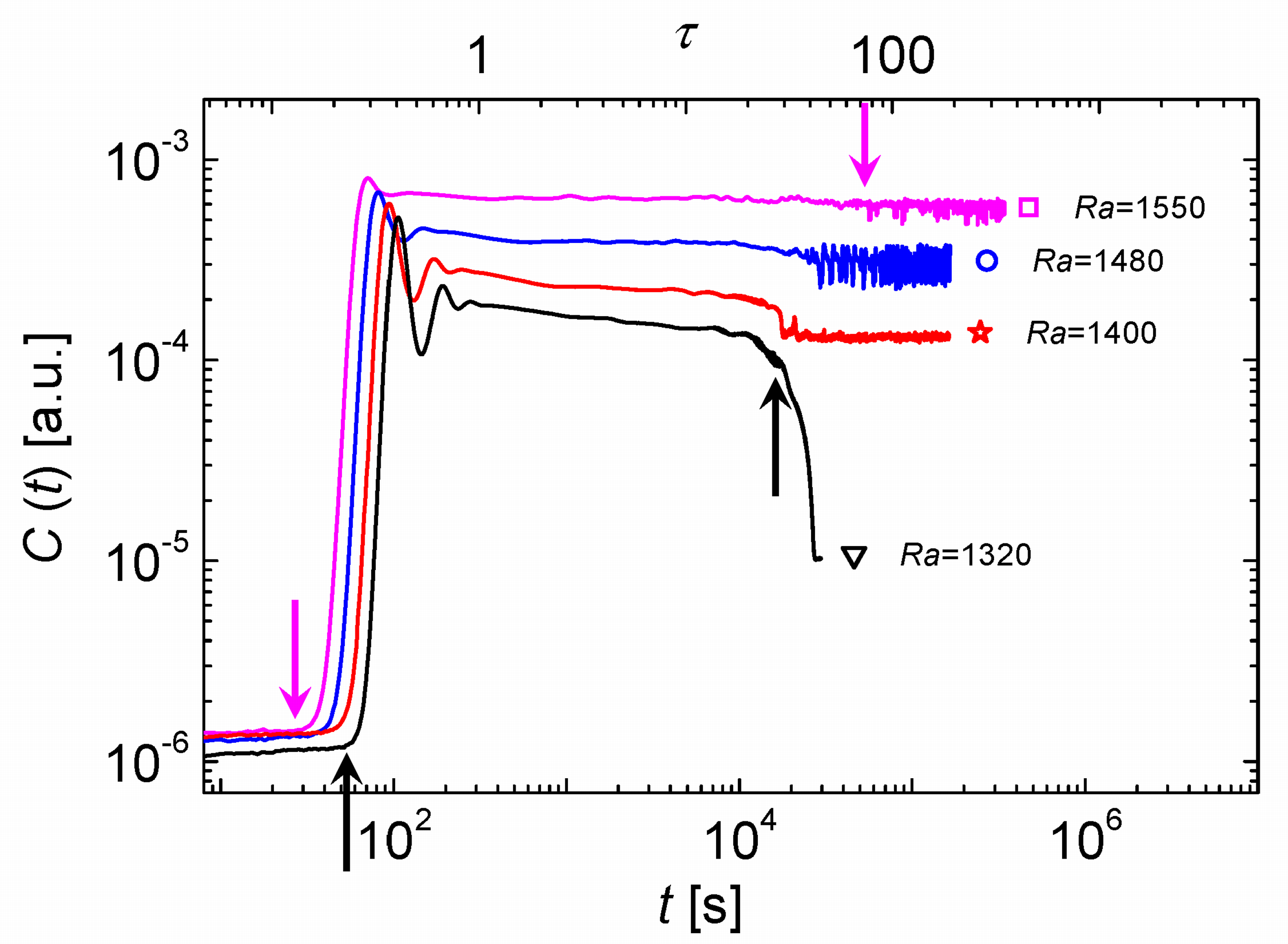}
\caption{(color online) Image contrast: contrast of shadowgraph image
sequences $C(t)$ plotted as a function of time (or dimensionless time relative to the diffusive time scale, top axis).
The layer of liquid is tilted at an angle of $24mrad$. The curves correspond to different imposed
temperature differences. Continuous lines stand for Rayleigh numbers: (from
top to bottom) $Ra$=1550, 1480, 1400, 1320. The left vertical arrows
indicate $t^{\ast }$ while the right ones indicate $\protect\tau _{LSF}$, as
detailed in the text.}
\label{fig phase diag}
\end{figure}
\newline
\indent Interestingly, up to the time $t_{LSF}$ the evolution of the
contrast under different experimental conditions is qualitatively similar.
After that time patterns becomes strongly affected by the Large Scale Flow
that determines a dramatic differentiation of the contrast. For an
inclination of $24mrad$, at $Ra=1320$ the drop of the contrast marks the
appearance of almost featureless patterns associated to the LSF (black line
in Fig. 3). The fact that the contrast does not drop back to its original
value is the signature of the presence of barely detectable fluctuations
induced by the shear motion. At $Ra=1400$ the balance between LSF and
solutal convection gives rise to a SH convective regime (red line in Fig. 3)
with a stable contrast, accompanied by small number fluctuations determined
by the entrance and exit of the columnar structures into the field of view.
At $Ra=1480$ the competition between LSF and solutal convection gives rise
to the transition between different patterns and the contrast exhibits
pronounced oscillations (blue line in Fig. 3). Finally, at $Ra=1550$ solutal
convection gives rise to square patterns, but the influence of the LSF still
determines a drifting of the patterns that gives rise to small number
fluctuations (magenta line in Fig. 3). \newline
\indent The presence of relaxation oscillations is a signature that the
onset of convection is determined by the destabilization of concentration
boundary layers (BLs). Basically, the BLs grow uniformly by diffusion until
they reach a critical thickness $\delta ^{\ast }$ beyond which convection
sets in. An estimate of the critical thickness of the boundary layers can be
obtained from the critical dimensionless latency time $\tau ^{\ast }=t^{\ast
}D/h^{2}$ \cite{delta*,howard66,degiorgio78,cerbino05}, where $t^{\ast }$ is
experimentally determined as the time at which the derivative of the image
contrast increases by a fixed amount (left arrows in Fig.3). Figure 4a-b
shows the dimensionless latency time $\tau ^{\ast }$ and critical BL
thickness $\delta ^{\ast }$, relative to $h$, plotted as a function of $Ra$.
Interestingly, the data in Fig. 4a-b fall onto the same curve independently
of the inclination angle. This confirms that the initial stages of pattern
formation are not significantly influenced by a marginal inclination of the
sample. 
\begin{figure}[t]
\center\includegraphics[width=8cm]{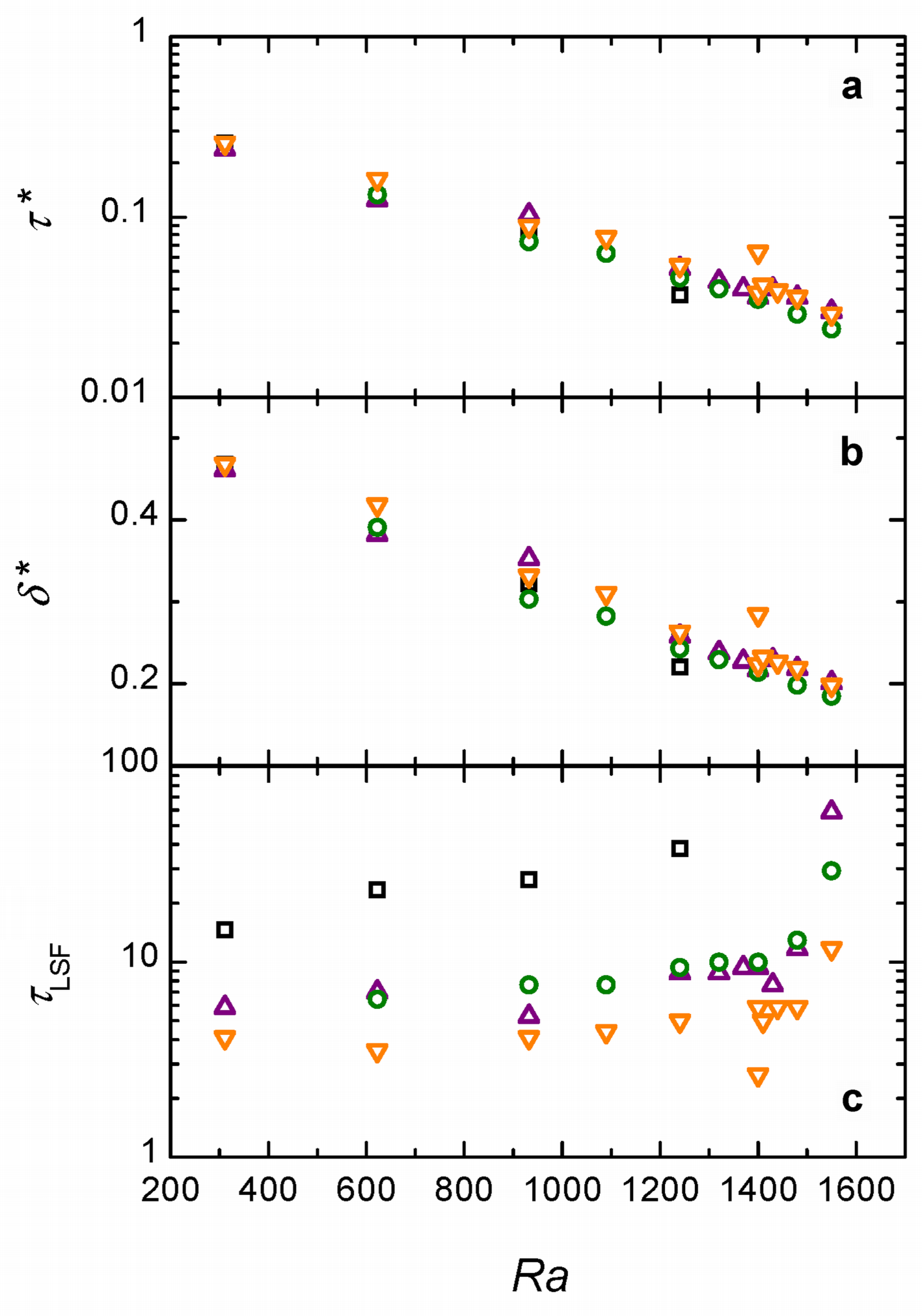}
\caption{(color online) Convection timescales and boundary layer thickness:
a, Dimensionless critical time $\protect\tau ^{\ast }$ for the onset of
convection. b, Critical thickness $\protect\delta ^{\ast }$ of the
dimensionless concentration boundary layer. c, Dimensionless time $\protect\tau$ for the
manifestation of the large-scale shear flow (LSF) related to the
inclination, $\protect\tau _{LSF}$. Different symbols stand for different
inclination angles: ($\square $) are for 6 mrad, ($\triangle $) for 18 mrad,
($\circ $) for 24 mrad and ($\triangledown $) for 36 mrad.}
\label{fig times}
\end{figure}
\newline
\indent After a time $t_{LSF}$ of the order of $10^{4}s$ (Fig. 3) the
patterns start to drift, thus marking the start of the influence of the
large scale shear flow induced by the inclination of the cell. Equivalently
the image contrast shows a drop (right arrows in Fig.3), which is used to
quantitatively identify $t_{LSF}$.  The dimensionless time $\tau _{LSF}=$
$t_{LSF}D/h^{2}$, which expresses \ $t_{LSF}$ relative to the diffusion time
across the cell height, is strongly affected by the inclination angle, as it
is apparent from the lack of overlap of the curves corresponding to
different inclination angles (Fig. 4c). For large Rayleigh numbers close to
the threshold $Ra_{c}$ the time $\tau _{LSF}$ shows signs of a divergence.
At such large $Ra$ a slowing down of the shear flow occurs, indicating the
dominance of the solutal columnar convective mass transfer over the shear
flow. \newline
\indent In the present work we limited our analysis to one liquid mixture
with a positive Soret coefficient; it remains an open question whether a
similar mechanism can be generally observed in other systems where a
temperature and composition stratification coexist. One interesting case is
the thermohaline circulation in oceans \cite{wunsch02, rahmstorf03}. Here
the cooling determined by wind at the surface of the ocean determines a
decrease of the temperature and an increase of the salinity through
evaporation, in the presence of a variable landscape at the bottom of the
ocean. The combined effect of these factors is one of the components leading
to the large-scale thermohaline circulation, a phenomenon that is still not
understood well, due to the simultaneous presence of other effects that
contribute to the oceanic currents. Our experimental model system may allow
to isolate and study independently one of the fundamental mechanisms driving
the thermohaline circulation process. In turn this will facilitate progress
towards a fundamental understanding of the interplay between large scale
flow and local columnar flows. \newline
\indent We thank Henri Bataller for critical reading of the manuscript, Georges Br\"ugger for experimental help and
Roberto Cerbino for early discussions. F.C. acknowledges financial support
from the European Union under FP7 PEOPLE Marie Curie Intra European
Fellowship, Contract No. IEF-251131, DyNeFI Project. This project has been
financially supported by the Swiss National Science Foundation (Project Nr.
132736). \newline
\indent Correspondence and requests for materials should be addressed to
F.C. (fabrizio.croccolo@univ-pau.fr)


\begin{thebibliography}{99}
\bibitem{delta*} The dimensionless critical thickness $\delta^*$ is the
relative thickness of the solutal boundary layers as they become instable.
As their growth is diffusive, their critical value can be demonstrated to
be: 

$\delta^*=\frac{1}{2}-\frac{4}{\pi^2}\sum_{n=0}^{\infty}\frac{1}{(2n+1)^2}%
exp[-\pi^2(2n+1)^2t^*] $. 
\end{thebibliography}
\end{document}